\documentclass{ifacconf}

\usepackage{graphicx}      
\usepackage{natbib}        
\usepackage{amssymb,amsmath}
\usepackage{color,soul}

\usepackage{epstopdf}

\UseRawInputEncoding
\usepackage{multicol}
\usepackage{subfigure}
\newtheorem{lem}{Lemma}
\newtheorem{theorem}{Theorem}

\newcommand{\norm}[1]{\left\lVert#1\right\rVert}
\DeclareMathAlphabet{\pazocal}{OMS}{zplm}{m}{n}

\begin{document}
\begin{frontmatter}

\title{Experiment design for impulse response identification with signal matrix models \thanksref{footnoteinfo}} 


\thanks[footnoteinfo]{This work is supported by the Swiss National Science Foundation under grant
no. $200021\_178890$.}

\author[First]{Andrea Iannelli,}
\author[First]{Mingzhou Yin,}
\author[First]{Roy S. Smith}

\address[First]{Automatic Control Lab, ETH, Z\"{u}rich 8092, Switzerland \\ (e-mail: iannelli/myin/rsmith@control.ee.ethz.ch).}

\begin{abstract}                
This paper formulates an input design approach for truncated infinite impulse response identification in the context of implicit model representations recently used as basis for data-driven simulation and control approaches. Precisely, the considered model consists of a linear combination of the columns of a data (or signal) matrix. An optimal combination for the case of noisy data was recently proposed using a maximum likelihood approach, and the objective here is to optimize the input entries of the data matrix such that the mean-square error matrix of the estimate is minimized. A least-norm problem is derived in terms of the optimality criteria typically considered in the experiment design literature.
Numerical results showcase the improved estimation fit achieved with the optimized input.
\end{abstract}


\end{frontmatter}

\section{Introduction}

In system identification, experiment design, originally developed in statistics \citep{Draper1996}, is concerned with determining optimal experimental conditions that maximize the accuracy of the identified model according to a pre-defined quality measure and subject to constraints defining admissible conditions. In the case of open-loop experiments, the conditions primarily refer to the input data used to excite the system \citep{Mehra_survey_1974}. The most established experiment design methods are framed in the context of parametric identification \citep{Goodwin_book}. Here, the problem can be recast as the minimization of various measures of the parameter covariance matrix (i.e. the inverse of the Fisher information matrix). The problem can then be efficiently solved in the frequency-domain by parameterizing the input by its spectrum and enforcing power constraints \citep{Jansson_TAC05}. Less often considered is the problem of input design in other settings, e.g. in subspace identification the only cases that have been investigated are conditions on the input such that the estimates are consistent and the correct model order can be estimated \citep{Chui_IJC_05}. %

The increasing interest in data-driven methods has put more emphasis on non-parametric representations of dynamical systems, whereby predictions and decisions are often made directly from the available data (without constructing an explicit model of the plant). The objective of this paper is to propose an input design approach for data-driven prediction problems
by leveraging established ideas from experiment design.
More precisely, truncated infinite impulse response (IIR) identification with the non-parametric estimator recently proposed in \citep{Ming_SMM_Arxiv} under the name of signal matrix model (SMM) is considered. The SMM approach is inspired by results from behavioural system theory \citep{Willems_2005}, which provide implicit representations of linear time-invariant systems by means of data (or signal) matrices made of input-output trajectories. They have proven very powerful both for simulation and control \citep{Markovsky_2008}, but are only valid for deterministic (noise-free) trajectories. In \citep{Ming_SMM_Arxiv} a solution for noisy data was proposed in the form of an approximated maximum likelihood estimation (MLE) problem, which also provides a statistical characterization of the estimation error. Leveraging this framework, the input design problem is formulated by minimizing a measure of the mean-square error matrix of the estimated truncated IIR. Notably, for the classic A-, D-, and E- optimality criteria, it is shown that this is equivalent to solving the same least-norm problem.

Recent works have considered input design for finite impulse response (FIR) models from an information theoretic perspective \citep{Fujimoto_Auto_2018,MuChen_Auto_2018} using the regularized least-squares approach as the estimator. In this paper, finiteness of the impulse response is not assumed, and for this
the use of the implicit SMM estimator is instrumental. Novel aspects also include the fact that the optimal input depends on the true system (this is typical in experiment design problems but uncommon for FIR estimation unless regularization is used) and the challenge of adding the MLE constraint. The problem is solved in the time-domain, i.e. the input trajectory is directly optimized over, and thus amplitude constraints, in addition to energy or power constraints, can be easily considered. Numerical results confirm the benefit of using the optimal inputs when comparing the estimates' accuracy with input signals commonly used. Robustness properties of the algorithm, together with the effect of the noise on the quality of the data-fit, are also investigated by Monte Carlo simulation.

%
%
%
%
%



\section{Background material}

\subsection{Notation}

The symbol $\pazocal{N}(\mu,\Sigma)$ indicates a Gaussian distribution with mean $\mu$ and covariance $\Sigma$. The expectation and the covariance of a random vector $x$ are denoted by $\mathbb{E}(x)$ and $\text{cov}(x)$ respectively. For a matrix $X$, the operation of stacking its columns in a single vector is denoted by $\text{vec}(X)$; $X^\dagger$ indicates the Moore-Penrose pseudoinverse; $(X)_{i,j}$ denotes the $(i,j)$-th entry of $X$. The symbol $I_n$ denotes the identity matrix of dimension $n$.
¨
Given a signal $x:\mathbb{Z}\to \mathbb{R}^n$, its trajectory from $k$ to $k+N-1$ is indicated as $(x_t)_{t=k}^{k+N-1}$, and in vector form as $\mathbf{x}=\text{col}(x_k,\dots,x_{k+N-1})$ by stacking its entries row-wise.
The block Hankel matrix $\mathcal{H}_{l}((x_t)_{t=i}^{j})$ with depth $l$ associated with $(x_t)_{t=i}^{j}$ is:
\begin{equation*}
\mathcal{H}_{l}((x_t)_{t=i}^{j}):=\begin{bmatrix}
    x_{i} & x_{i+1} & \cdots & x_{j-l+1} \\
    x_{i+1}& x_{i+2} & \cdots & x_{j-l+2} \\
    \vdots & \vdots &  & \vdots \\
    x_{i+l-1} &  x_{i+l}  & \cdots   & x_{j}\\
    \end{bmatrix}.
\end{equation*}

\subsection{Impulse response estimation and least-squares approach}\label{FIR_LS}

Consider the infinite impulse response (IIR) representation of a single-input single-output discrete-time system:
\begin{equation}\label{IIR_def}
y_t = \sum^\infty_{i=0} h_i u_{t-i},
\end{equation}
where $y_t$ and $u_t$ are respectively output and input of the system at timestep $t$, and $(h_i)_{i=0}^\infty$ denotes the IIR.
The system identification problem considered here is the estimation of the truncated IIR of order $n$, i.e. of the first $n$ coefficients $(h_i)_{i=0}^{n-1}$, using input $(u_i^d)_{i=1}^{N}$ and (noisy) output $(y_i^d)_{i=1}^{N}$ trajectories of length $N$.

A classic approach to approximately solve this problem is to seek a finite impulse response (FIR) representation of order $n$ of the system \citep{LjungBook2}.
The FIR identification problem can then be postulated as a linear regression by vectorizing a finite expansion of Eq. (\ref{IIR_def}):
\begin{equation}\label{FIR_vect}
    \underbrace{\begin{bmatrix}y^d_0\\y^d_1\\\vdots\\y^d_{N-1}\end{bmatrix}}_{y_N}=\underbrace{
    \begin{bmatrix}u^d_0&u^d_{-1}&\cdots&u^d_{1-n}\\u^d_1&u^d_{0}&\cdots&u^d_{2-n}\\
    \vdots&\vdots&\ddots&\vdots\\u^d_{N-1}&u^d_{N-2}&\cdots&u^d_{N-n}\end{bmatrix}}_{\Phi_N}
    \underbrace{\begin{bmatrix}h_0\\h_1\\\vdots\\h_{n-1}\end{bmatrix}}_{h}.
\end{equation}
When the Toeplitz regressor matrix $\Phi_N$ has full column rank, an estimate for $h$ can be obtained from the unique least-squares (LS) solution of (\ref{FIR_vect}). Under the assumption that the output $y_t$ is contaminated by i.i.d. noise with distribution $\pazocal{N}(0,\sigma^2)$, the least-squares estimates have statistics:
\begin{equation}
    \mathbb{E}(h) = \left(\Phi_N^{\top}\Phi_N\right)^{-1}\Phi_N^{\top} y_N,\, \; \text{cov}(h) = \sigma^2\left(\Phi_N^{\top}\Phi_N\right)^{-1}.
    \label{eqn:LS}
\end{equation}

The mean-squared error of the estimates (\ref{eqn:LS}) crucially depends on the values of $n$ and $N$. It is known that a larger model flexibility gained by increasing $n$ comes at the cost of a larger variance, especially when $N$ is of the same order of $n$. Regularized least-squares problems have been widely investigated in the context of FIR estimation to address this bias-variance trade-off \citep{Chen_Auto12,Pillonetto_2010}, and recently in \citep{Fujimoto_Auto_2018,MuChen_Auto_2018} the input design problem for regularized LS identification of FIR models has been considered.

Even with regularization, there are two important aspects underlying the LS formulation. First, the past input trajectory $(u_i^d)_{i=1-n}^{-1}$ is required. If this is not available, either the past inputs are assumed zero (and, when this does not hold, bias is introduced), or the first $n-1$ data points are discarded (non-windowed case), with detrimental consequences on the data efficiency.
Second, Eq. (\ref{FIR_vect}) solves in general a different problem than the truncated IIR estimation originally stated and of interest here, unless $h_i\approx 0$ for all $i\geq n$. Satisfaction of this condition depends on the dominant poles of the system, and for lightly damped plants large values of $n$ are required.

\subsection{The signal matrix model}

To overcome the aforementioned limitations, it is used here as truncated IIR estimator an alternative approach recently proposed in \citep{Ming_SMM_Arxiv}.
The signal matrix model (SMM) is an implicit model representation  
inspired by results from behavioural system theory \citep{Willems_2005} and providing favourable statistical properties in the case of noisy-data. A compounded version of the underpinning results, instrumental to present the SMM framework, is reported in the following theorem.

\begin{theorem}\citep{Willems_2005,Markovsky_2008}
Consider a controllable system with McMillan degree $n_x$ and an input-output trajectory $(u_i^d, y_i^d)_{i=0}^{N-1}$ where $\mathcal{H}_{L+n_x}((u_t^d)_{t=0}^{N-1})$ has full row rank (input persistently exciting of order $L + n_x$). Then, $(u_i,y_i)_{i=0}^{L-1}$ is an input-output trajectory of this system iff there exists $g$ such that:
    \begin{equation}\label{Willems_1}
\text{col}(u_0,\dots,u_{L-1}, y_0,\dots,y_{L-1})
=\begin{bmatrix}\mathcal{H}_{L}((u_t^d)_{t=0}^{N-1})\\\mathcal{H}_{L}((y_t^d)_{t=0}^{N-1})\end{bmatrix}g,
\end{equation}
where the Hankel matrices have $M=N-L+1$ columns. Moreover, given $L_0\geq n_x$, the \emph{past} input-output trajectory $(u_i,y_i)_{i=-L_0}^{-1}$ uniquely determines the initial condition $x_0 \in \mathbb{R}^{n_x}$ of the underlying state. Therefore, by denoting $L'=L-L_0$, it holds from (\ref{Willems_1}) that $(y_i)_{i=0}^{L'-1}$  (i.e. $\mathbf{y}$) is the unique output trajectory of the system with past trajectory $(u_i,y_i)_{i=-L_0}^{-1}$ (denoted by $\mathbf{u}_{\text{ini}}, \mathbf{y}_{\text{ini}}$) and input trajectory $(u_i)_{i=0}^{L'-1}$  (i.e. $\mathbf{u}$), iff there exists $g$ such that:
       \begin{equation}\label{eqn:fund}
    \begin{bmatrix} \mathbf{u}_{\text{ini}} \\ \mathbf{u} \\ \hline    \mathbf{y}_{\text{ini}} \\ \mathbf{y} \end{bmatrix}
    =\begin{bmatrix}U\\ \hline Y \end{bmatrix}g
    =\begin{bmatrix}U_p\\U_f\\ \hline Y_p\\Y_f \end{bmatrix}g
    =\begin{bmatrix}\mathcal{H}_{L_0+L'}((u_t^d)_{t=0}^{N-1})\\ \hline \mathcal{H}_{L_0+L'}((y_t^d)_{t=0}^{N-1})\end{bmatrix}g,
    \end{equation}
    where $U$ and $Y$ are partitioned according to \emph{past} and \emph{future} trajectories.
\label{thm:1}
\end{theorem}
Given the initial condition ($\mathbf{u}_{\text{ini}}, \mathbf{y}_{\text{ini}}$) and a persistently exciting input $(u_i^d)_{i=0}^{N-1}$ of a sufficiently high order, Theorem \ref{thm:1} allows the response of the system $\mathbf{y}$ to any input signal $\mathbf{u}$ to be exactly predicted (or simulated) by finding a value of $g$ which satisfies (\ref{eqn:fund}). Compactly, this implicit model can be expressed as:
\begin{equation}\label{eqn:ddmodel}
    \mathbf{y}=f(\mathbf{u};\mathbf{u}_{\text{ini}},\mathbf{y}_{\text{ini}}, U_p,U_f,Y_p,Y_f).
\end{equation}
When any of the signals and signal matrices on the right hand side of (\ref{eqn:ddmodel}) are subject to noise, Theorem \ref{thm:1} does not hold and it is not clear how to best determine $g$.
A statistical approach was adopted in \citep{Ming_SMM_Arxiv}, yielding a maximum likelihood estimate (MLE) of $g$.

We consider here the case where $(y_i^d)_{i=0}^{N-1}$
is built from measurements contaminated by i.i.d. Gaussian noise:
\begin{equation}
    y_i^d = y_i^{d,0}+w_i^d,\; \; (w_i^d)_{i=0}^{N-1}\sim \pazocal{N}(0,\sigma^2 I_N).
\end{equation}
Note that $\mathbf{y}_{\text{ini}}$ is assumed noise-free here, since the interest is on simulation problems where the initial condition is set by the analyst.

\begin{lem}\label{lem:MLE}\citep{Ming_SMM_Arxiv}
The value of $g$ that maximizes the conditional probability of observing
$\hat{\mathbf{y}}=Yg-\left[\begin{smallmatrix}\mathbf{y}_{\text{ini}}\\\mathbf{0}\end{smallmatrix}\right]$
given $g$ is obtained by solving:
\begin{equation}
\underset{g\in\pazocal{G}}{\text{min}}\
    \text{logdet}(\Sigma_y(g))+\begin{bmatrix}Y_pg-\mathbf{y}_{\text{ini}}\\\mathbf{0}\end{bmatrix}^{\top}\Sigma_y^{-1}(g)\begin{bmatrix}Y_pg-\mathbf{y}_{\text{ini}}\\\mathbf{0}\end{bmatrix},
\label{eqn:opt0}
\end{equation}
where $\pazocal{G}$ is the parameter space:
\begin{equation}
    \pazocal{G} = \left\{g\in \mathbb{R}^M\left|\begin{bmatrix}
    U_p\\U_f
    \end{bmatrix}g=\begin{bmatrix}
    \mathbf{u}_{\text{ini}}\\\mathbf{u}
    \end{bmatrix}\right.\right\},
\end{equation}
and
\begin{equation}    \label{eqn:stats}
    \left(\Sigma_y\right)_{i,j} = \left( \text{cov}(\hat{\mathbf{y}}|g)\right)_{i,j}= \sigma^2\sum_{k=1}^{M-|i-j|}g_k g_{k+|i-j|}.
\end{equation}
\end{lem}
Once $g$ is determined from (\ref{eqn:opt0}), the output $\mathbf{y}$ for the selected input and initial conditions is given by $\mathbf{y}=Y_f g$. To efficiently solve (\ref{eqn:opt0}), two approximations are employed in \citep{Ming_SMM_Arxiv}: $\Sigma_y$ is replaced by its diagonal relaxation $\bar{\Sigma}_y$, which is obtained by setting to zero all the off-diagonal terms, namely $\bar{\Sigma}_y=\sigma^2 \norm{g}_2^2 I_{L}$; the nonlinear terms appearing in (\ref{eqn:opt0}) are approximated such that the objective function is quadratic in $g$. With these simplifications, problem (\ref{eqn:opt0}) has the closed-form solution:
\begin{equation}\label{g_analytic}
\begin{aligned}
  &  g_{\text{SMM}}= F^{-1}U^{\top}(U F^{-1}U^{\top})^{-1}\tilde{\mathbf{u}}, \\
\text{where} \quad & \tilde{\mathbf{u}}:=\text{col}(\mathbf{u}_{\text{ini}},\mathbf{u}), \; \; F:=L \sigma^2 I_{M} +Y_p^{\top}Y_p. \\
\end{aligned}
\end{equation}

The signal matrix model lends itself to the identification of the truncated IIR of order $n$ of a system. This is indeed the particular simulation problem defined by:
\begin{equation}
    \mathbf{u}_{\text{ini}}=\mathbf{0},\,\mathbf{y}_{\text{ini}}=\mathbf{0},\,\mathbf{u}=\text{col}(1,\mathbf{0}),\,L'=n.
    \label{eqn:ddimp}
\end{equation}
Then, the sought estimate can be written as:
\begin{equation}
h=\mathbf{y}=Y_f g_{\text{SMM}}. 
\label{eqn:impEST}
\end{equation}
This approach does not rely on the FIR assumption made in the LS method, and instead provides an estimate of the truncated IIR. Unlike the LS approach, the SMM estimator (\ref{g_analytic}-\ref{eqn:impEST}): is correct (i.e., when applied to noise-free data, it gives the true model); is unbiased for any $n$ under the assumptions of Theorem \ref{thm:1}; 
does not require input measurements prior to the experiment. 
In the context of deterministic trajectories (Theorem \ref{thm:1}), an analogous behavioural estimator for the truncated IIR was initially proposed in \citep{Markovsky_2005b}.

\section{Experiment design with the SMM}

The truncated IIR estimation via the SMM approach hinges on the computation of $g_{\text{SMM}}$. As shown analytically in Eq. (\ref{g_analytic}), the latter will be a function of the Hankel matrices $U$ and $Y$, which in turn are built up from system's trajectories. The objective of this work is to design an input trajectory $(u_i^d)_{i=0}^{N-1}$, so that the corresponding matrices $U$ and $Y$ endow $g_{\text{SMM}}$ with some \emph{desirable properties}.

\subsection{Information criterion for SMM}\label{information_SMM}

The most desired feature for $g_{\text{SMM}}$ is to provide an estimated impulse response model which is as close as possible to the true one. One possible way to measure the quality of the estimate is via the mean-square error (MSE) matrix \cite[Chapter~7]{LjungBook2} of $h$ or, equivalently, of $\mathbf{y}$, denoted here by $S$. The objective of obtaining an accurate model can then be recast as the minimization of some measure of this matrix.
In input design problems, typical measures are represented by the A-, D-, and E- optimality criteria, respectively $\textnormal{Tr}\left(S\right)$, $\textnormal{logdet}\left(S\right)$, and $\lambda_{\text{max}}\left(S\right)$ (where $\lambda_{\text{max}}(\cdot)$ denotes the largest singular value of a matrix).
The following result establishes the connection between these criteria and the SMM estimator.
\begin{lem}\label{MSE}
Provided that the input trajectory $(u_i^d)_{i=0}^{N-1}$ is persistently exciting of order $L + n_x$, then
\begin{equation}
\mathcal{J}_{\bullet}(S)=f_{\bullet}(\norm{g_{\text{SMM}}}_2^2),
\end{equation}
where $\mathcal{J}_{\bullet}$ is any of the three commonly used input design optimality criteria ($\bullet=A,D,E$) applied to the MSE matrix $S$, and $f_{\bullet}(\cdot)$ is a specific function, different for each case, but with the property that it is a monotonically increasing function of its argument.
\end{lem}
\begin{pf}
Under the assumption of input persistently exciting, $h$ is in the range of $Y_f$ and thus the estimator is unbiased. Thus, the MSE matrix $S$ coincides with the covariance $\text{cov}(\mathbf{y}|g_{\text{SMM}})=\bar{\Sigma}_y=\sigma^2 \norm{g_{\text{SMM}}}_2^2 I_L$. Therefore, the following hold:
\begin{equation}\label{MSE_J_f}
\begin{aligned}
\mathcal{J}_{A}(S)&=\textnormal{Tr}\left(\bar{\Sigma}_y\right)=L \sigma^2 \norm{g_{\text{SMM}}}_2^2,\\
\mathcal{J}_{D}(S)&=\textnormal{logdet}\left(\bar{\Sigma}_y\right)=L \textnormal{log}\left(\sigma^2 \norm{g_{\text{SMM}}}_2^2\right),\\
\mathcal{J}_{E}(S)&=\lambda_{\text{max}}\left(\bar{\Sigma}_y\right)=\sigma^2 \norm{g_{\text{SMM}}}_2^2,\\
\end{aligned}
\end{equation}
where $\lambda_{\text{max}}(\cdot)$ denotes the largest singular value of a matrix. The functions $\mathcal{J}_{\bullet}$ in (\ref{MSE_J_f}) are all monotonically increasing function of $\norm{g_{\text{SMM}}}_2^2$. \qed
\end{pf}
This result shows that minimizing standard measures of the MSE matrix $S$ for the approximate SMM problem leads to the problem of minimizing the squared Euclidean norm of $g_{\text{SMM}}$. It is also noted that, in the case of A-optimality, this would also hold true for the original full (and not diagonal) expression of the covariance $\Sigma_y$ (\ref{eqn:stats}).
Therefore, SMM estimators featuring vectors $g_{\text{SMM}}$ with small Euclidean norm exhibit favourable statistical properties.

An optimization problem can then be formulated to find the input sequence such that the Euclidean norm of the solution $g_{\text{SMM}}$ of problem (\ref{g_analytic}) is minimized. Specializing $\tilde{\mathbf{u}}$ to the identification of the impulse response case yields:
\begin{subequations}\label{ED_goal}
\begin{align}
    \underset{g,(u_i^d)_{i=0}^{N-1}}{\text{min}} & \norm{g}_2^2, \label{ED_goal_1}\\
    \text{s.t.} & \; \; g=g_{\text{SMM}} = F^{-1}U^{\top}(U F^{-1}U^{\top})^{-1} \begin{bmatrix}\mathbf{0}\\ 1 \\ \mathbf{0}\end{bmatrix}, \label{ED_goal_2}\\
& (u_i^d)_{i=0}^{N-1} \in \mathcal{U}. \label{ED_goal_3}
\end{align}
\end{subequations}
where $\mathcal{U}$ defines input constraints (e.g. amplitude bounds), and $U$ and $F$ are defined in (\ref{eqn:fund}) and (\ref{g_analytic}), respectively.

\subsection{Proposed problem solution}\label{optimization_SMM}

Solving problem (\ref{ED_goal}) presents two major challenges.

First, the matrix $F$ (\ref{g_analytic}) depends on $Y_p$, and thus on the output of the system to the input trajectory, which in turn is the identification objective of the design. Recalling the definition of $Y_p$ (\ref{eqn:fund}) and by virtue of its Hankel structure, $F$ depends on the sequence $(y_i^d)_{i=0}^{N-L'-1}$.
This has to do with the formulation of the identification problem with the implicit representation in Eq. \ref{eqn:ddmodel}. The MSE matrix of the estimator depends explicitly only on $g$, which in turn is a function of both signal matrices $U$ and $Y$ via (\ref{g_analytic}).

While this feature appears to be peculiar to the SMM model formulation, the fact that the optimal input depends on the true system that one intends to identify is a known limitation of many experiment design approaches \citep{Goodwin_book}. In this spirit, the solution proposed here consists of adopting a \emph{baseline} model to approximately estimate the sequence $(y_i^d)_{i=0}^{N-L'-1}$ as a function of the input trajectory (precisely, its first $N-L'$ data points). The sought linear map between input and output can then be approximated as
\begin{equation}\label{baseline_vect}
    \begin{bmatrix}\tilde{y}^d_0\\\tilde{y}^d_1\\\vdots\\\tilde{y}^d_{N-L'-1}\end{bmatrix} \approx \underbrace{
    \begin{bmatrix}h^b_0& 0 &\cdots& 0 \\
    h^b_1 & h^b_{0}&\ddots&0\\
    \vdots&\ddots&\ddots&\vdots\\h^b_{N-L'-1}&h^b_{N-L'-2}&\cdots&h^b_{0}\end{bmatrix}}_{H^b}
    \begin{bmatrix}u^d_0\\u^d_1\\\vdots\\u^d_{N-L'-1}\end{bmatrix},
\end{equation}
where the square Toeplitz matrix $H^b$ consists of the coefficients $(h_i^b)_{i=0}^{n_b}$, with $n_b-1\leq N-L'$, of the baseline model. When $n_b-1 < N-L'$, the last $N-L'-n_b-1$ coefficients are set to zero and $H^b$ will have zero diagonals in the corresponding lower left part. The baseline model reflects prior knowledge of the system and can be obtained either from first principles or from an identification based on a previous experiment. As shown in \citep{Chen_Auto12} in the context of regularized LS estimation, the use of baseline models can be seen from a Bayesian viewpoint as putting a prior on the mean of the impulse response. By virtue of (\ref{baseline_vect}), $Y_p$ will be approximated in the design problem by $\tilde{Y}_p(u^d,H^b)$, which is a matrix function with Hankel structure linearly dependent on the optimized input trajectory $(u_i^d)_{i=0}^{N-L'-1}$ for a fixed baseline model $H^b$:
\begin{equation}\label{Yp_baseline}
\begin{aligned}
\tilde{Y}_p(u^d,H^b)&=\begin{bmatrix}
    \tilde{y}^d_0 & \tilde{y}^d_1 & \cdots & \tilde{y}^d_{N-L_0-L'+1} \\
    \tilde{y}^d_1 & \tilde{y}^d_2 & \cdots & \tilde{y}^d_{N-L_0-L'+2} \\
    \vdots & \vdots &  & \vdots \\
    \tilde{y}^d_{L_0-1} &  \tilde{y}^d_{L_0}  & \cdots   & \tilde{y}^d_{N-L'-1}\\
    \end{bmatrix},\\
    &=\mathcal{H}_{L_0}((\tilde{y}_i^d)_{i=0}^{N-L'-1}).\\
\end{aligned}
\end{equation}

The second challenge is that constraint (\ref{ED_goal_2}) is highly nonlinear in the input trajectory. This constraint enforces that the vector $g$, whose Euclidean norm has to be minimized, solves the MLE problem and is thus an SMM estimator. The solution proposed here is to replace (\ref{ED_goal_2}) by the KKT conditions associated with the approximated version of the MLE problem (\ref{eqn:opt0}), which are an equivalent set of constraints but more amenable to numerical optimization. Specifically, the MLE problem (\ref{eqn:opt0}) obtained by replacing the original covariance $\Sigma_y$ with its diagonal relaxation $\bar{\Sigma}_y$ is:
\begin{equation}
\begin{matrix}
    &
 \underset{g}{\text{min}} &    \text{logdet}(\bar{\Sigma}_y(g))+\frac{\norm{Y_p g}_2^2}{\sigma^2\norm{g}_2^2}, \\
    &\text{s. t.} & \begin{bmatrix}
    U_p\\U_f
    \end{bmatrix}g=\begin{bmatrix}
    \mathbf{0} \\ \text{col}(1,\mathbf{0})
    \end{bmatrix}.
\end{matrix}
\label{eqn:optsqp}
\end{equation}
The KKT conditions are then derived by using the same approximations proposed in \citep{Ming_SMM_Arxiv}. In particular: the first term is approximated by a first-order expansion around the point $g_0$, i.e. $ \text{logdet}(\bar{\Sigma}_y(g)) \approx \text{logdet}(\bar{\Sigma}_{y}(g_0))+\text{tr}\left(\bar{\Sigma}_{y}(g_0)^{-1}\left(\bar{\Sigma}_{y}(g)-\bar{\Sigma}_{y}(g_0)\right)\right)$; the bilinearity in the second term is resolved by taking a zero-order approximation of the denominator around the point $g_0$. As a result, the expression for the objective function of (\ref{eqn:optsqp}), aside for constant terms, is approximated by:
\begin{equation*}
\begin{aligned}
    &\text{tr}\left(\bar{\Sigma}_{y}(g_0)^{-1} \bar{\Sigma}_{y}(g)\right)+\frac{1}{\sigma^2 \norm{g_0}_2^2} \norm{Y_p g}_2^2,\\
    =& L\frac{\norm{g}_2^2}{\norm{g_0}_2^2}+\frac{\norm{Y_p g}_2^2}{\sigma^2 \norm{g_0}_2^2}  = \frac{1}{\sigma^2\norm{g_0}_2^2}\left( L \sigma^2 \norm{g}_2^2 + \norm{Y_p g}_2^2 \right). \\
\end{aligned}
\end{equation*}
The Lagrangian for the optimization problem is then:
\begin{equation*}
\mathcal{L}(g, \nu)= L \sigma^2 \norm{g}_2^2 + \norm{Y_p g}_2^2+2\nu^{\top}\left(Ug-\tilde{\mathbf{u}}\right),
\end{equation*}
where $\nu \in \mathbb{R}^{L}$ are the Lagrangian multipliers and $\tilde{\mathbf{u}}=\text{col}(\mathbf{0},1,\mathbf{0})$. The KKT condition for the problem can be obtained by setting to zero the partial derivatives of $\mathcal{L}$ with respect to $g$ and $\nu$:
\begin{equation*}
\begin{aligned}
\frac{\partial \mathcal{L}}{\partial g} &= 2 \left( L \sigma^2 g+ Y_p^{\top}\left( Y_p g \right)+ U^{\top} \nu \right)=0,\\
\frac{\partial \mathcal{L}}{\partial \nu} &= Ug-\tilde{\mathbf{u}}=0,\\
\end{aligned}
\end{equation*}
This provides a suitable set of constraints replacing (\ref{ED_goal_2}).

The proposed experiment design problem can thus be written as
\begin{subequations}\label{ED_prog}
\begin{align}
&    \underset{g,(u_i^d)_{i=0}^{N-1},\nu}{\text{min}}  \norm{g}_2^2, \label{ED_prog_1}\\
    \text{s.t.} & \; \;
    \begin{bmatrix} L \sigma^2 I_M + \tilde{Y}_p(u^d,H^b)^{\top} \tilde{Y}_p(u^d,H^b)& U^{\top}\\ U & 0\end{bmatrix} \begin{bmatrix}g \\ \nu \end{bmatrix}=\begin{bmatrix} \mathbf{0} \\ \tilde{\mathbf{u}} \end{bmatrix}, \label{ED_prog_2}\\
& (u_i^d)_{i=0}^{N-1} \in \mathcal{U}, \label{ED_prog_3}
\end{align}
\end{subequations}
where $\tilde{Y}_p$ (\ref{Yp_baseline}) is used to approximate the unknown $Y_p$ as a linear function of the optimized input trajectory for a chosen baseline model. For $\mathcal{U}$, two typical constraints encountered in input design problems are considered here, namely energy and magnitude constraints:
\begin{equation}
\begin{aligned}
\mathcal{U}^{\text{En}} & = \left\{(u_i^d)_{i=0}^{N-1} \in \mathbb{R}^N | \sum_{i=0}^{N-1}(u_i^d)^2 \leq E_0 N \right\},\\
\mathcal{U}^{\text{Mag}}  & = \left\{(u_i^d)_{i=0}^{N-1} \in \mathbb{R}^N | \underline{u} \leq u_i^d \leq \overline{u}, \forall i \right\}.\\
\end{aligned}
\end{equation}
Program (\ref{ED_prog}) has a convex cost and convex constraint (\ref{ED_prog_3}), but constraint (\ref{ED_prog_2}), even though its expression has been significantly simplified with respect to the equivalent (\ref{ED_goal_2}), is still nonlinear. The interior point solver IPOPT \citep{IPOPT} is employed here. 


\section{Simulated experimental results}
Consider the following fourth-order system given in the $z$-transform domain:
\begin{equation}
    G(z) = \dfrac{k(z^3+0.5z)}{z^4-2.2z^3+2.42z^2-1.87z+0.7225},
    \label{eqn:sys1}
\end{equation}
which was originally studied in \citep{Pillonetto_2010} with $k=1$ (a truncated IIR is depicted in the bottom right plot of Fig. 3 of the reference). Here we take $k=0.1159$ to make $G$ of unitary $\mathcal{H}_2$ norm. This system has a relatively slow impulse response and it was used in \citep{Ming_SMM_Arxiv} to show the higher accuracy of the signal matrix model's truncated IIR estimates compared to LS-based FIR estimates, particularly with respect to higher robustness against truncation errors. Input data were generated therein using i.i.d. Gaussian signals with zero mean and unitary variance.
The goal here is to investigate the benefit of using the proposed experiment design approach to form the signal matrices employed in the SMM model.
Only SMM estimates will be considered throughout, since this allows one to isolate the effect of the input (focus of this work) from that of the algorithm (with the advantages of SMM with respect to LS being already investigated in \citep{Ming_SMM_Arxiv}).

The following measure of fit \citep{Chen_Auto12} is used to judge the accuracy of the $n$ estimated IIR coefficients:
\begin{equation*}
    W=100 \left(1-\left[\frac{\sum_{i=0}^{n-1}(h_i-\hat{h}_i)^2}{\sum_{i=0}^{n-1}(h_i-\bar{h})^2}\right]^{1/2}\right),
\end{equation*}
where $h_i$ are the true IIR coefficients, $\hat{h}_i$ are the estimated coefficients, and $\bar{h}$ is the mean of the true coefficients. The parameters used in the simulations are, unless otherwise specified,
$
    N=63,L_0=8,n=L'=13,\sigma^2=0.01, E_0=1, \overline{u}=-\underline{u}=\sqrt{E_0}
$. 
Note that $N\geq 2(L+n_x)-1$ and $L_0 \geq n_x$ as required by Theorem \ref{thm:1}.
The baseline model (with $n_b-1=L'$) is obtained with SMM using data collected in a prior experiment conducted using i.i.d. Gaussian inputs of length $N$ (having total energy $E_0 N$). Monte Carlo simulations with 200 different realizations of $(w_i^d)_{i=0}^{N-1}$ are carried out to analyze robustness to noise. Note in this regard that the signal-to-noise ratio (SNR) is given by $\frac{E_0}{\sigma^2}$.

\begin{figure}[ht]
\begin{center}
\subfigure[Energy constraint $\mathcal{U}^{\text{En}}$.]{%
\includegraphics[width=0.8\columnwidth]{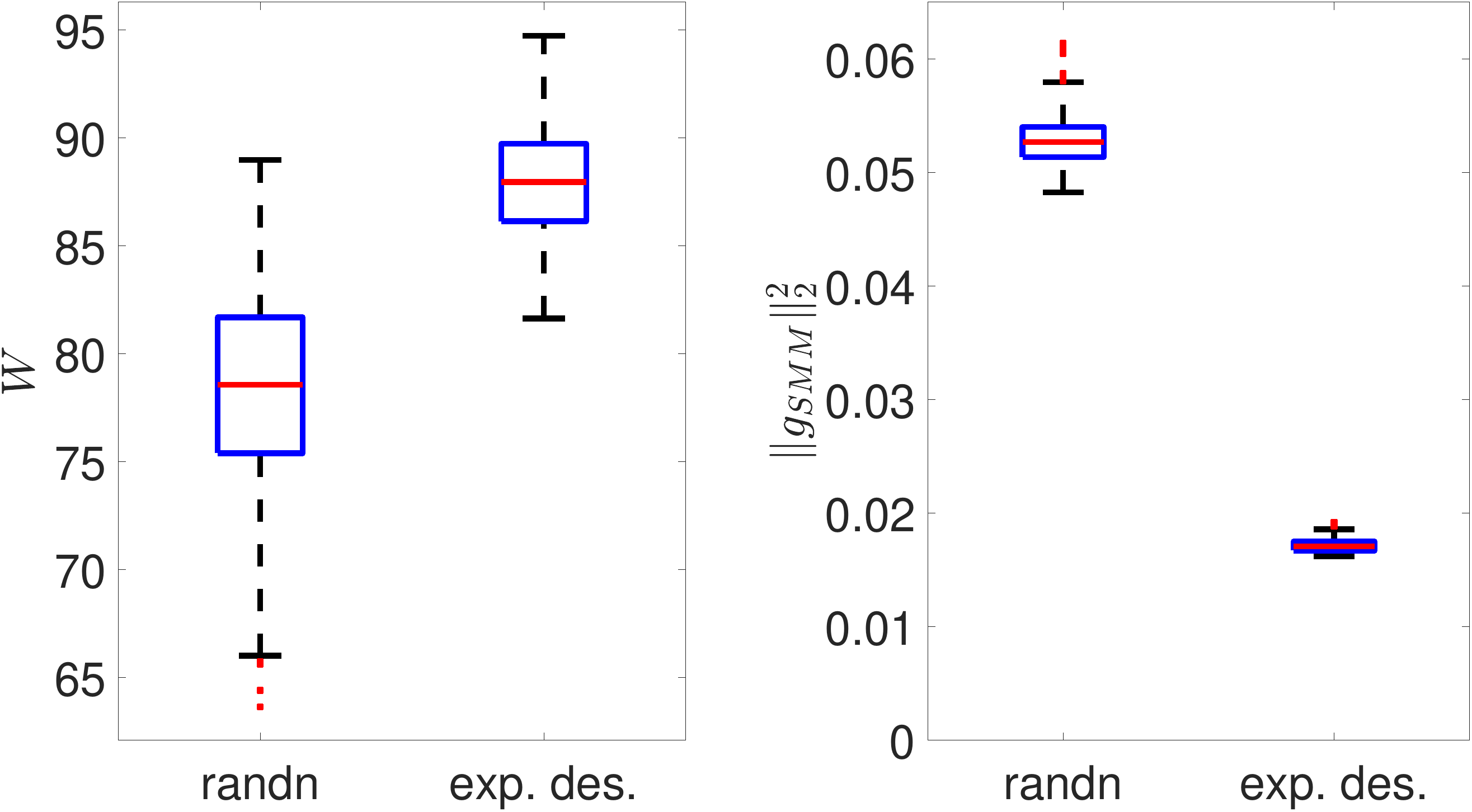}    
\label{Fig_1_En}}
\subfigure[Magnitude constraint $\mathcal{U}^{\text{Mag}}$.]{%
\includegraphics[width=0.8\columnwidth]{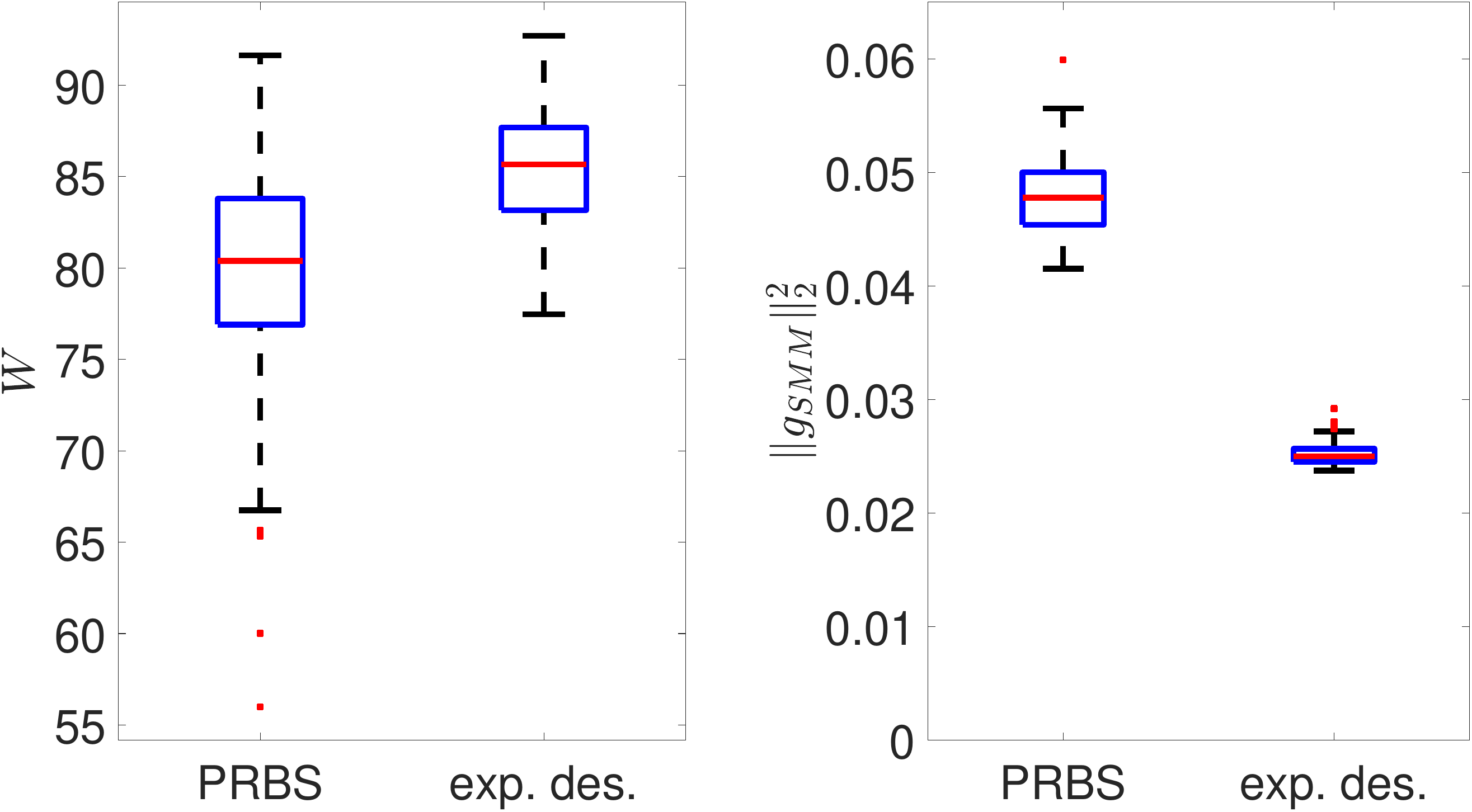}    
\label{Fig_1_Mag}}
\caption{Box plots for fit and $\norm{g_{\text{SMM}}}_2^2$: comparison between standard vs. experiment design inputs.}
\label{Fig_1}
\end{center}
\end{figure}

In the first test, we optimize input sequences $(u_i^d)_{i=0}^{N-1}$ with program (\ref{ED_prog}) for energy (Fig. \ref{Fig_1_En}) and magnitude constraints (Fig. \ref{Fig_1_Mag}). Fig. \ref{Fig_1} shows box plots for the measure of the fit $W$ and the squared Euclidean norm of the vector $g_{\text{SMM}}$ (\ref{g_analytic}) for the experiment design input (\emph{exp. des.}) and a standard alternative method. For constraint $\mathcal{U}^{\text{En}}$, this consists of a zero-mean i.i.d. Gaussian signal scaled in order to have energy $E_0 N$ (denoted by \emph{randn}), whereas for constraint $\mathcal{U}^{\text{Mag}}$ it is a pseudorandom binary sequence (PRBS) in the range $[ \underline{u}, \overline{u} ]$ (denoted by \emph{PRBS}). The reason for these choices is that these are, for the respective constraints, the optimal input for classic FIR estimation problems and thus are widely used. Recall that the SNR for these experiments is $\frac{E_0}{\sigma^2}=100$.

\begin{figure}[h!]%
\begin{center}
\includegraphics[width=0.82\columnwidth]{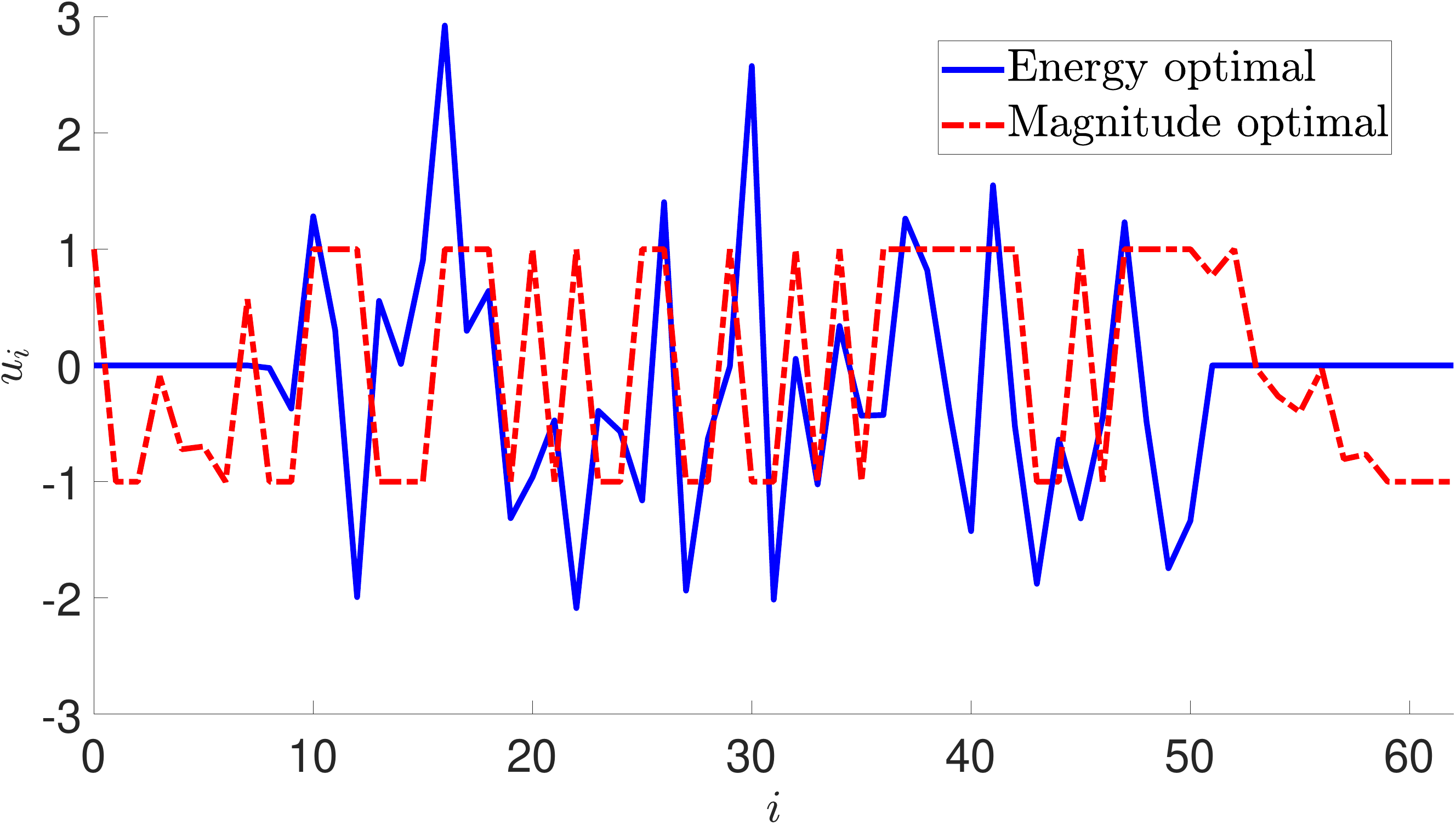}
\caption{Optimal inputs for SMM estimation for energy and magnitude constrained cases.}
\label{Fig_1_input}
\end{center}
\end{figure}

In both scenarios it is observed that using inputs designed with the proposed approach improves the fit of the estimated models, in terms of median fit and also dispersion. The right plots give insights into the reason for the better performance. It can indeed be seen that the vector $g_{\text{SMM}}$ computed by Eq. (\ref{g_analytic}) and used to estimate the IIR with Eq. (\ref{eqn:impEST}) has a much lower squared Euclidean norm, and hence a smaller covariance $\bar{\Sigma}_y$, when optimized inputs are used. This shows that the information metric defined in Lemma \ref{MSE} indeed has a direct impact on the accuracy of the SMM estimates, and that program (\ref{ED_prog}) effectively provides input sequences which make the norm of $g_{\text{SMM}}$ smaller. It is also observed that there is a much smaller dispersion of $\norm{g_{\text{SMM}}}_2^2$ when \emph{exp. des.} input are used.
Figure \ref{Fig_1_input} shows the optimized input signals for the cases of energy constraint (solid line) and magnitude constraint (dash-dotted line).

Interestingly, neither of the experiment design signals resemble the standard alternatives, respectively an i.i.d. sequence for energy and a (pseudo)random binary signal for magnitude constraint. In the energy case, the optimal signal has the first $L_0$ and the last $L'-1$ entries equal to 0. As for the magnitude case, the signal is not binary and as a result has less energy than the competitor PRBS considered in Fig. \ref{Fig_1_Mag} (but obtaining nonetheless better performance).

\begin{figure}[h!]%
\begin{center}
\includegraphics[width=0.82\columnwidth]{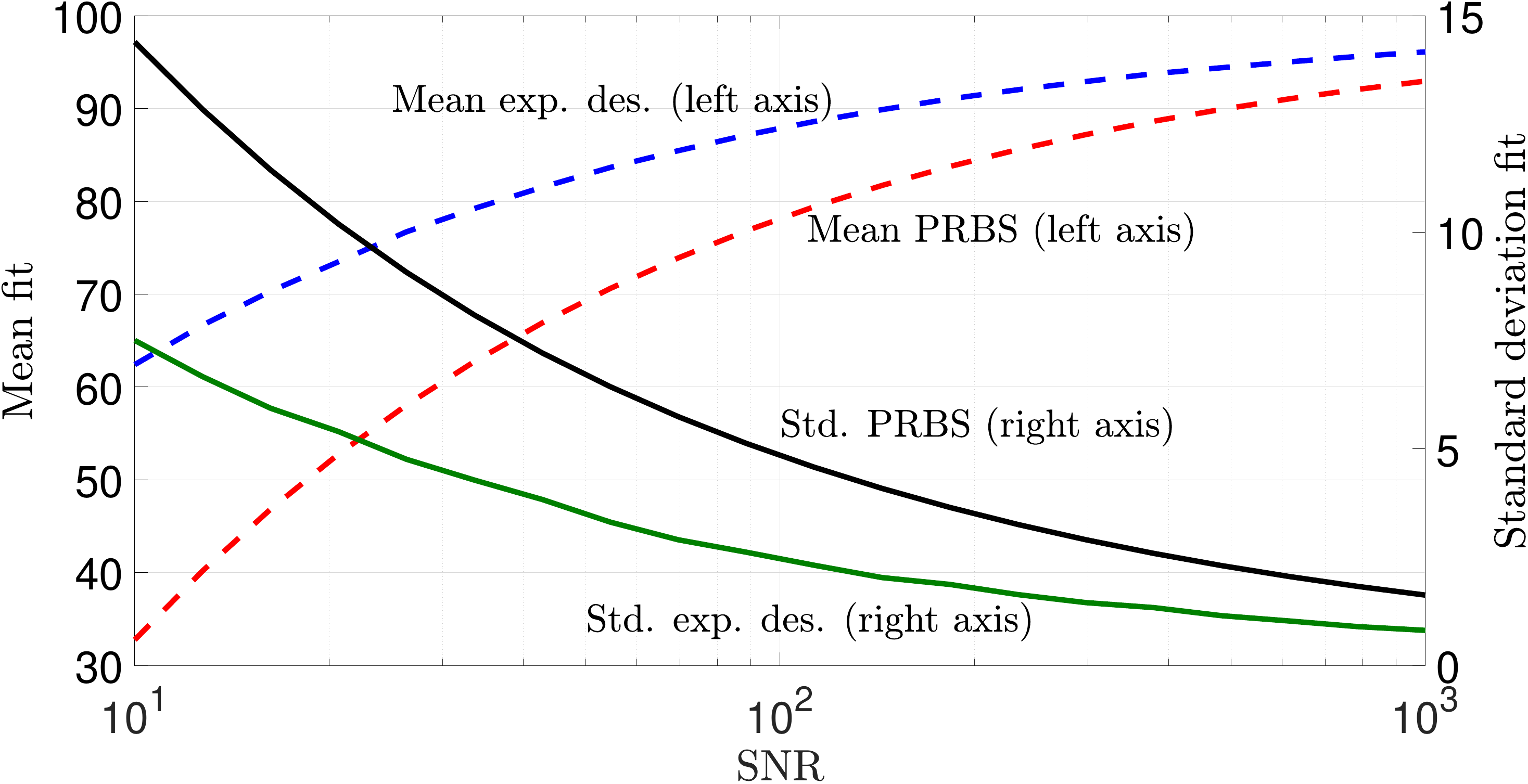}
\caption{Effect of SNR: comparison between i.i.d. input and optimized input (energy constraint case).}
\label{Fig_2_SNR}
\end{center}
\end{figure}

Next, we examine how the improvement in the fit achieved with the optimized input sequences changes with the SNR ratio of the experiments. Figure \ref{Fig_2_SNR} shows mean (left axis) and standard deviation (right axis) of the fit $W$ as $\sigma^2$ is decreased from 0.1 ($\frac{E_0}{\sigma^2}=10$) to 0.001 ($\frac{E_0}{\sigma^2}=1000$).

As expected, both estimates improve (the fit has larger mean and smaller variance) as the SNR increases. The estimates obtained with the optimized input always feature a markedly higher performance, and this improvement is more pronounced in the range of low SNR.

\begin{figure}[h!]%
\begin{center}
\includegraphics[width=0.82\columnwidth]{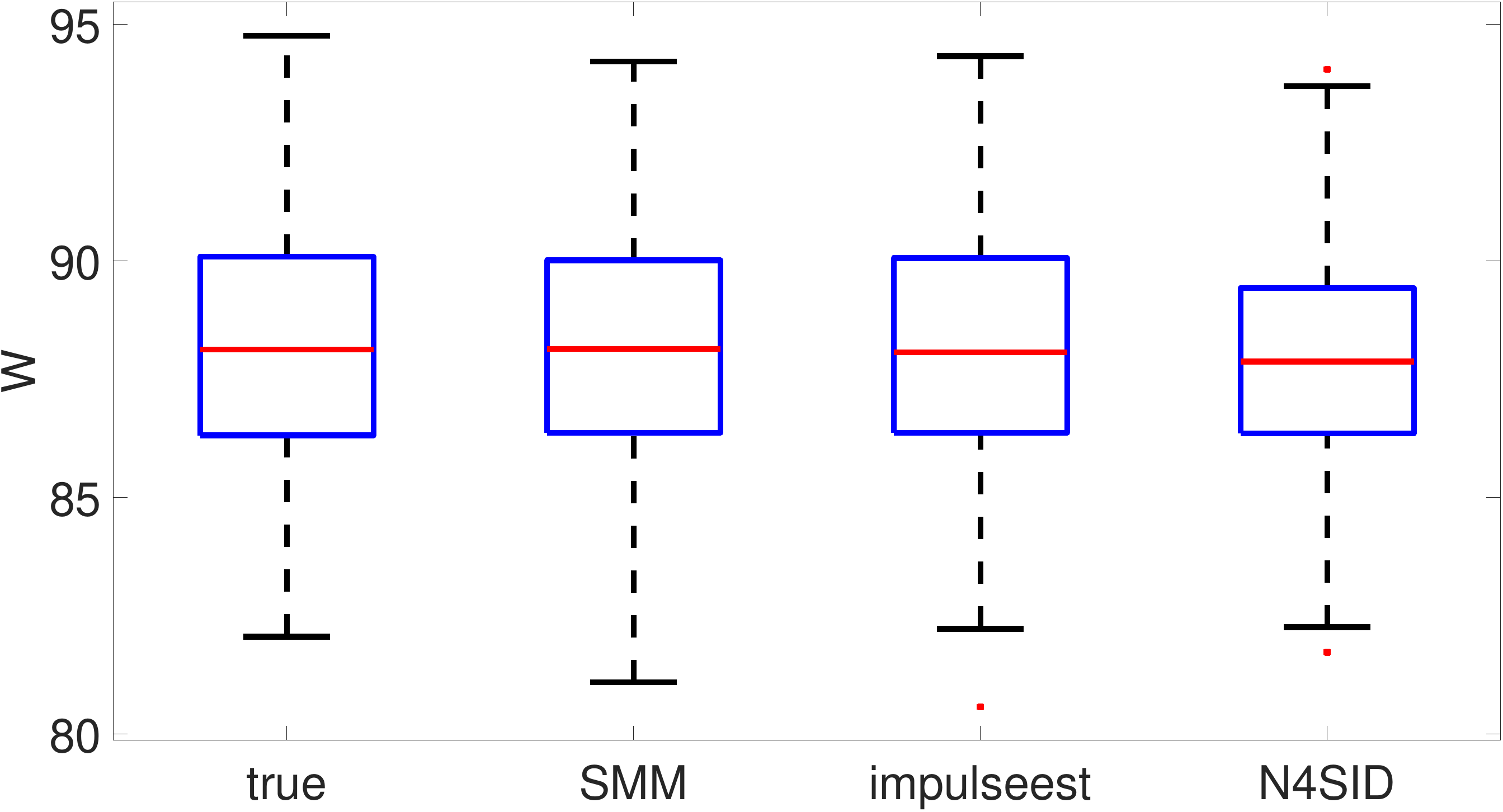}
\caption{Effect of different baseline models on the SMM estimate's fit obtained with optimized input.}
\label{Fig_3_base}
\end{center}
\end{figure}

Robustness of the optimized inputs to the employed baseline model $H^b$ is investigated by comparing the fit of SMM estimates computed with optimized input obtained using different baseline models. The experiment used to identify the baseline model consists as before of
i.i.d. input sequence of length $N=64$ and $E_0=1$, but now the output is contaminated with i.i.d. noise with covariance $\sigma^2=0.1$, thus the SNR of the experiment providing the baseline model has value 10. Results are shown in Fig. \ref{Fig_3_base} where the labels in the $x$ axis refer to the method with which $H^b$ has been obtained: in \emph{true} by the true system; in \emph{SMM} by an SMM estimate (as done in the previously shown results); in \emph{impulseest} by the homonymous MATLAB functions (default options); in \emph{N4SID} by realizing the state-space model identified with the homonymous MATLAB functions (default options and automatic determination of system order). Note that, due to the low SNR, the fit $W$ of the truncated IIR models used to build $H^b$ is poor: 34.4 for \emph{SMM}; 8.1 for \emph{impulseest}; 55.7 for \emph{N4SID}.

The results show that the accuracy of the estimates is almost unaffected by the choice of baseline model. Indeed, the fit is almost identical across the four baseline models $H^b$ which represent a markedly different estimate of the true system (represented by \emph{true}).

\section{Conclusions}
An input design formulation for truncated IIR estimation using implicit model representations via signal matrices has been presented. Leveraging a recently proposed statistical characterization of these data-driven estimates, a program is proposed to optimize input sequences such that the mean-square error matrix of the estimate is small. In fact, it is shown that the proposed least-norm problem solves the classic optimality criteria considered in the literature, but here in a non-parametric setting. The results show quantitative advantages of using the optimized inputs compared to commonly used input choices. 

\bibliography{SYSID21_exp}             


\end{document}